\documentclass[aps,prc,reprint,showpacs,superscriptaddress]{revtex4-1}

\usepackage{graphicx}
\usepackage{amsmath}
\usepackage{subfigure}
\usepackage{amssymb}
\usepackage{hyperref}

\begin{document}

\title{Penning-trap mass spectrometry of highly charged, neutron-rich Rb and Sr isotopes in the vicinity of $A\approx100$}

\date{\today}

\author{V.~V.~Simon} 
\email[Corresponding author: ]{vsimon@triumf.ca (V. V. Simon)}
\affiliation{TRIUMF, 4004 Wesbrook Mall, Vancouver, BC, V6T 2A3, Canada}
\affiliation{Department of Physics and Astronomy, University of Heidelberg, Philosophenweg 12, 69120 Heidelberg, Germany}
\affiliation{Max Planck Institute for Nuclear Physics, Saupfercheckweg 1, 69117 Heidelberg, Germany}

\author{T.~Brunner}
\affiliation{TRIUMF, 4004 Wesbrook Mall, Vancouver, BC, V6T 2A3, Canada}
\affiliation{Physik Department E12, Technische Universit\"at M\"unchen, 85748 Garching, Germany} 

\author{U.~Chowdhury}
\affiliation{TRIUMF, 4004 Wesbrook Mall, Vancouver, BC, V6T 2A3, Canada}
\affiliation{Department of Physics and Astronomy, University of Manitoba, Winnipeg, MB, R3T 2N2, Canada}

\author{B.~Eberhardt}
\affiliation{TRIUMF, 4004 Wesbrook Mall, Vancouver, BC, V6T 2A3, Canada}

\author{S.~Ettenauer}
\affiliation{TRIUMF, 4004 Wesbrook Mall, Vancouver, BC, V6T 2A3, Canada}
\affiliation{Department of Physics and Astronomy, University of British Columbia, Vancouver, BC, V6T 1Z1, Canada}

\author{A.~T.~Gallant}
\affiliation{TRIUMF, 4004 Wesbrook Mall, Vancouver, BC, V6T 2A3, Canada}
\affiliation{Department of Physics and Astronomy, University of British Columbia, Vancouver, BC, V6T 1Z1, Canada}

\author{E.~Man\'e}
\affiliation{TRIUMF, 4004 Wesbrook Mall, Vancouver, BC, V6T 2A3, Canada}

\author{M.~C.~Simon}
\affiliation{TRIUMF, 4004 Wesbrook Mall, Vancouver, BC, V6T 2A3, Canada}

\author{P.~Delheij}
\affiliation{TRIUMF, 4004 Wesbrook Mall, Vancouver, BC, V6T 2A3, Canada}

\author{M.~R.~Pearson}
\affiliation{TRIUMF, 4004 Wesbrook Mall, Vancouver, BC, V6T 2A3, Canada}

\author{G.~Audi}
\affiliation{CSNSM-IN2P3-Universit\'{e} de Paris Sud, Orsay, France}

\author{G.~Gwinner}
\affiliation{Department of Physics and Astronomy, University of Manitoba, Winnipeg, MB, R3T 2N2, Canada}

\author{D.~Lunney}
\affiliation{CSNSM-IN2P3-Universit\'{e} de Paris Sud, Orsay, France}

\author{H.~Schatz}
\affiliation{National Superconducting Cyclotron Laboratory, Michigan State University, E. Lansing, MI, 48824, USA}

\author{J.~Dilling}
\affiliation{TRIUMF, 4004 Wesbrook Mall, Vancouver, BC, V6T 2A3, Canada}
\affiliation{Department of Physics and Astronomy, University of British Columbia, Vancouver, BC, V6T 1Z1, Canada}

\begin{abstract}

The neutron-rich mass region around $A\approx100$ presents challenges for modeling the astrophysical $r$-process because of rapid shape transitions. We report on mass measurements using the TITAN Penning trap at TRIUMF-ISAC to attain more reliable theoretical predictions of $r$-process nucleosynthesis paths in this region. A new approach using highly charged ($q=15+$) ions has been applied which considerably saves measurement time and preserves accuracy.  New mass measurements of
neutron-rich $^{94,97,98}$Rb and $^{94,97-99}$Sr have uncertainties of less than 4~keV and show deviations of up to 11$\sigma$ to previous measurements. An analysis using a parameterized $r$-process model is performed and shows that mass uncertainties for the $A=90$ abundance region are eliminated.

\end{abstract}


\maketitle

\section{Introduction}
The atomic mass, and from it the derived nuclear binding energy, is a key property of the nuclear many-body system. The binding energy reflects all interactions among the constituents and provides important information for a broad variety of studies in nuclear and atomic physics \cite{Lunney2003, Blaum2006} including tests of the standard model of particle physics \cite{Towner2010}, nuclear structure studies and tests of theory and models (e.g. \cite{Bender2008}), and nuclear astrophysics \cite{Langanke2003}, including nucleosynthesis pathways. 
In addition, the conserved vector current (CVC) hypothesis and the unitarity of the Cabbibo-Kobayashi-Maskawa (CKM) quark mixing matrix (\cite{Towner2010} and references therein) are probed by determining $V_{ud}$ from super-allowed $\beta$ emitters. One of the key inputs is the $Q$-value obtained from the mass difference between the mother and daughter of the decay.
Nuclear structure variations from nucleus to nucleus can be sensitively probed by differences in nuclear masses and thus provide insight to sudden or gradual changes in nuclear structure.
(See discussion in \cite{Naimi2010}, recently confirmed by $\gamma$-ray spectroscopy by \cite{Albers2012}.) 
Here we report on new measurements of masses of neutron-rich nuclei in the $A\approx100$ region.

\begin{figure*}
    \begin{center} 
	\includegraphics[width=17.2cm] {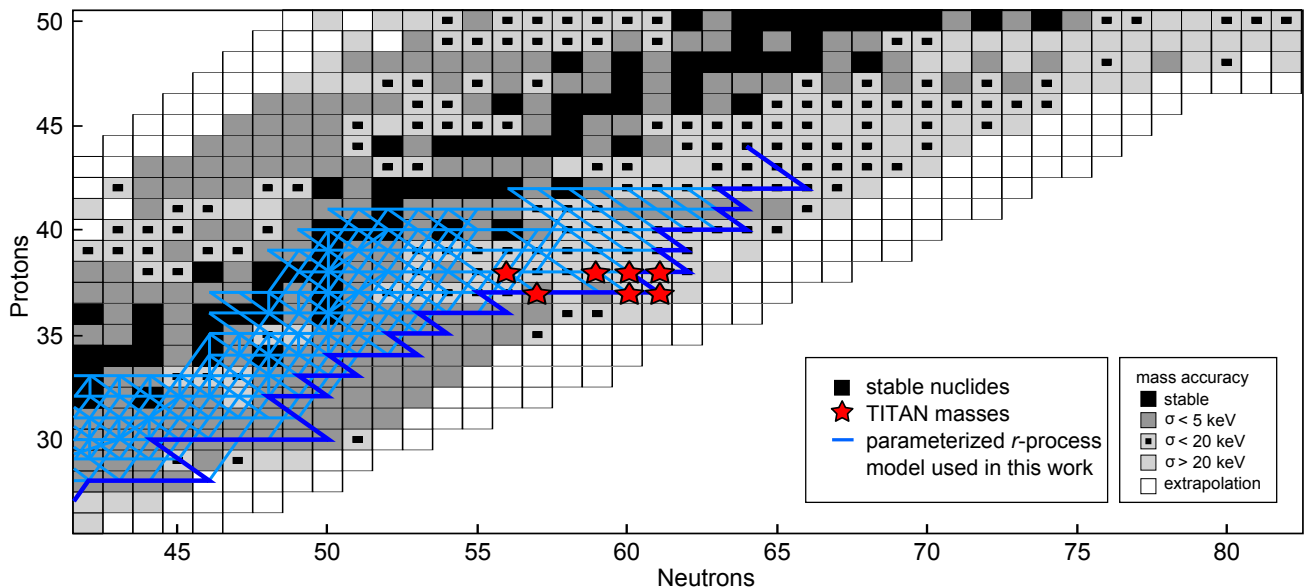} 
    \end{center}
    \caption{\label{fig:nuclidcharta}(color online) A section of the nuclide chart. The neutron-rich nuclei studied in this article are indicated by red stars. 
    The mass accuracy is displayed as well as the calculated time-integrated net-reaction-flows for a single $S=100$ component (blue) of the parameterized, high entropy wind inspired, $r$-process model used in this work. Flows above a relative final abundance of $10^{-5}$ are denoted by light blue lines 
indicating the complex interplay of charged-particle and neutron-induced reactions in this mass region; dark blue lines mark the outer boundary of the reaction flows for clarity.}
\end{figure*}

The synthesis of about half of the heavy elements beyond germanium ($Z=32$) proceeds in nature via the rapid neutron-capture process ($r$-process) \cite{Arnould2008}. In the most common $r$-process models the neutron capture reactions proceed until an equilibrium between neutron capture and photo disintegration, (n,$\gamma$) $\longleftrightarrow$ ($\gamma$,n), is established, driving the $r$-process path to nuclei with neutron separation energies of approximately 2 to 3~MeV \cite{Langanke2003}.

Testing $r$-process models against abundance observations requires reliable nuclear physics that translates a set of evolving astrophysical conditions into a characteristic nuclear abundance pattern. Nuclear input data of importance to $r$-process models are nuclear masses, $\beta$-decay half-lives and branching ratios for neutron emission. Fission rates and fission fragment distributions, neutrino interaction rates, and charged particle fusion rates also play a role \cite{Langanke2010}. 

In models characterized by an (n,$\gamma$) $\longleftrightarrow$ ($\gamma$,n) equilibrium, the reaction path for a given set of astrophysical conditions is governed by nuclear masses. However, many of the nuclei in the $r$-process are inaccessible experimentally. Hence, nuclear astrophysics calculations rely heavily on theoretical mass predictions, which are performed by models with parameters that are adjusted to known masses \cite{Lunney2003}. Experimental mass data on neutron-rich nuclei are therefore not only needed as direct input into $r$-process model calculations, but also to improve mass models and reduce the need for extrapolations. 
Fig.~\ref{fig:nuclidcharta} shows the reaction flows of the parameterized, fully dynamical $r$-process model that is used in this work, following Freiburghaus et al. \cite{Freiburghaus1999} and Hosmer et al. \cite{Hosmer2010}. The model is based on an adiabatic expansion as it might be encountered in high entropy neutrino driven winds in core collapse supernovae. The path is characterized by a complex network of charged particle and neutron induced reactions and their inverse. More details can be found later in the text. The model passes through the mass region covered by our experiment ($A\approx100$), which helps in reducing the challenge for theoretical mass models.

Additional motivation for this work stems from the desire to better understand the nuclear structure, in particular shell closure effects at $N=50$, the subshell closure at $N=54$, and a possible onset of large deformation for nuclei with $N\geq60 $ \cite{Andreyev2000}. 
Several theoretical investigations of the nuclear structure in the $N\approx60$ region have been carried out (\cite{Kumar1985, Rodriguez-Guzman2010, Rodriguez-Guzman2011} and references therein). Within a self-consistent Hartree-Fock-Bogoliubov (HFB) formalism \cite{Kumar1985}, the nearly-spherical shape for $N<59$ Sr, Zr, and Mo isotopes is predicted to develop into a strongly deformed prolate shape above $N>60$. In the so-called D1S-Gogny energy-density-functional framework (\cite{Rodriguez-Guzman2011} and references therein), in which one-quasiparticle configurations are employed, a self-consistent mean-field approximation indicates nuclear deformation and shape co-existence. These calculations predict a sharp oblate-to-prolate transition in the Rb, Sr, and Zr isotopes and triaxiality in the Mo isotopes.

Empirical evidence supports rapidly changing behavior in $N\approx60$ nuclei \cite{Hager2006, Rahaman2007, Hager2007, Hakala2011, Campbell2002, Naimi2010, Albers2012}.
For example, nuclear deformation can manifest itself in the two-neutron separation energy $S_{2n}$, the difference in energy in isotopes separated by two neutrons. Typically, $S_{2n}$ decreases smoothly with the neutron number $N$, and a change in slope may indicate the onset of deformation or (sub)shell closures. This signature has already been observed 
in the Rb isotopes \cite{Epherre1979}, and  
in the Sr and Zr isotopic chains around $N\approx60$ \cite{Hager2006}; however, neutron-rich Kr isotopes do not present any nuclear quantum phase transition \cite{Naimi2010}. In this work, we extend the investigation of nuclear deformation around $N\approx60$ to neutron-rich Rb and Sr isotopes via Penning-trap mass spectrometry.

Previous Penning trap mass measurements in this region \cite{Rahaman2007, Hager2006} differ from the atomic mass evaluation (AME03) \cite{Audi2003b} by up to 300~keV and up to 11$\sigma$. Therefore, an independent confirmation was desired. 
To that end, precise and accurate mass measurements on radioactive, short-lived isotopes have been performed at TRIUMF's Ion Trap for Atomic and Nuclear science setup (TITAN) \cite{Dilling2003, Dilling2006} at the radioactive beam facility for Isotope Separator and ACcelerator (ISAC) \cite{Dombsky2000}. 
A unique 
feature of TITAN over other Penning traps at radioactive ion beam facilities is the ability to charge-breed ions prior to the mass measurement. The advantage of high charge states is the improved precision, which scales directly with the charge state $q$. 
Nuclei with short life times can benefit from this approach because it enables high precision with short observation times in the Penning trap. This has recently been demonstrated for $^{74}$Rb \cite{Ettenauer2011} ($T_{1/2}=65$~ms). As long as the production yields of the nuclei are high enough to counteract additional losses introduced by the charge breeding process, the full precision gain of a factor of q can be exploited. At lower yields the actual gain is reduced but will stay above unity provided that the overall charge breeding efficiency is greater than $1/q^2$. Alternatively, by boosting the precision the required uncertainties can be reached in shorter times which enables the mapping of larger parts of the nuclear chart during the limited measurement times at rare beam facilities. 
The experimental time available to measure the seven masses in this work was only $\approx$40~hours which was suitable to achieve a precision that is relevant for nuclear astrophysics as described later in this publication. A conservative approach concerning systematic uncertainties was taken into account.

\section{Experimental Setup}

\begin{figure}
    \begin{center}
	\includegraphics[width=8.6cm]{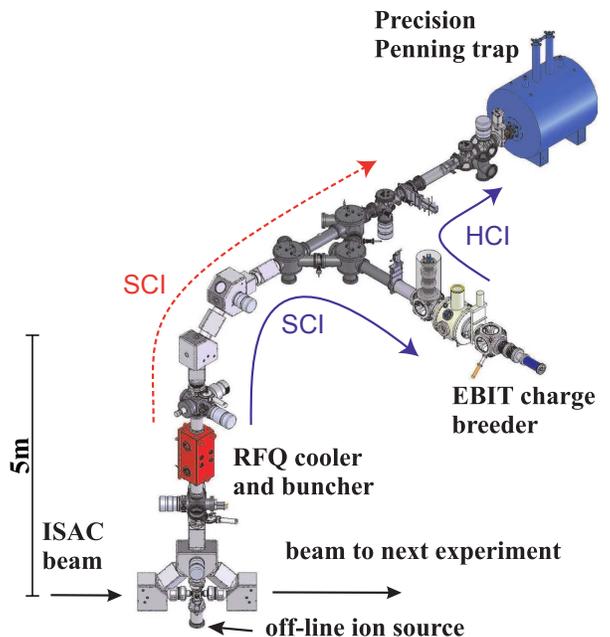} 
    \end{center}
    \caption{\label{fig:TITAN}(color online) The TITAN facility for high-precision atomic mass measurements is presently composed of three ion traps. The singly charged ion (SCI) beam from ISAC passes through an RFQ (radio-frequency quadrupole) where the ions are thermalized in a helium buffer gas, bunched, and then sent straight to the precision Penning trap for mass measurements (MPET), indicated by the dashed red arrow. Alternatively, with solid blue arrows the pathway for mass measurement of highlycharged ions (HCI) is shown. The SCI from ISAC are sent to the EBIT (electron-beam ion trap), charge-bred, and transported to MPET.}
\end{figure}

TITAN is a multi-trap setup located at the ISAC \cite{Dombsky2000} facility at TRIUMF. 
The radioactive isotopes are produced by bombarding an ISOL-type target with 500~MeV protons from the cyclotron. 
For the first time, a UC$_{x}$ target was used at TRIUMF and neutron-rich isotopes were produced for this experiment by bombarding it with a proton current of 2~$\mu$A \cite{Dombsky2012}. 
The ion beam was generated using a surface-ionization source and extracted at 20~keV, transported through the ISAC mass separator with a resolving power $m/\delta m\approx3000$ and delivered to the TITAN facility. TITAN, shown schematically in Fig.~\ref{fig:TITAN}, presently consists of three traps: a buffer-gas-filled radiofrequency quadrupole (RFQ) trap \cite{Brunner2012} for cooling and bunching, an electron-beam ion trap (EBIT) \cite{Lapierre2010} for charge breeding, and a hyperbolic Penning trap (MPET) for high-precision mass measurements 
on short-lived nuclei with a precision of down to $\delta m / m\approx10^{-8}$ \cite{Brodeur2009a}. 

\begin{figure}
    \begin{center}
	\includegraphics[width=8.6cm]{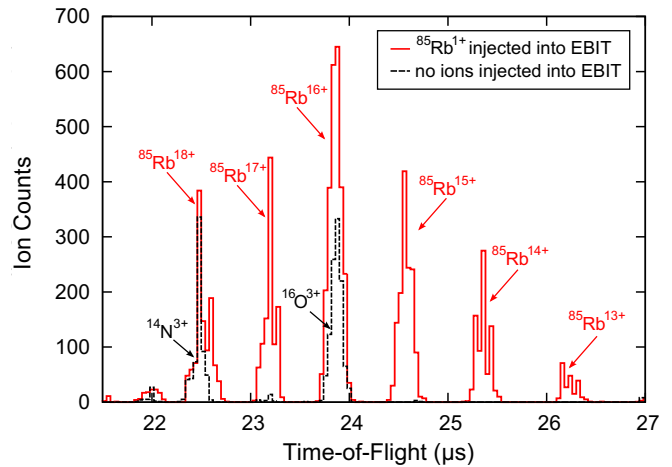} 
    \end{center}
    \caption{\label{fig:chargebreeding}(color online) Time-of-flight spectra of charge-bred ions extracted from the EBIT. A charge-breeding time of 197~ms, a magnetic field of 3~T, an electron beam current and energy of 30~mA and 2.5~keV, respectively, and an extraction time of 800~ns were applied to obtain this spectra. The data represent 500 ion bunches. The red line shows the case of injection of $^{85}$Rb$^{1+}$ into the EBIT. The black spectrum represents background gas in the EBIT.}
\end{figure}

While previous work at TITAN focused on mass measurements of singly charged ions (SCI), in particular light neutron-rich isotopes and isotopes with neutron halos \cite{Smith2008, Ryjkov2008, Ringle2009, Ettenauer2010, Brodeur2011, Brodeur2012a}, the present work used charge-bred ions (see Fig.~\ref{fig:TITAN}). 
Mass measurements of short lived, highly charged ions (HCI) were recently performed at TITAN \cite{Ettenauer2011, Lapierre2012} for the first time and here we extend this to neutron-rich Rb, Sr isotopes. 
In this scheme the ions are first cooled and bunched in the RFQ, and then sent to the EBIT (see Fig.~\ref{fig:TITAN}) where they are recaptured. 
An energetic electron beam (up to 70~keV) removes bound electrons of the initially singly charged ions through impact ionization.
In this experiment the ions were charge-bred for $T_{breed}\approx80$~ms using an electron beam current and energy of 30~mA and 2.5~keV, respectively. 
Ions were then extracted by opening the trap barrier for 800~ns, generating a short pulse. This leads to a sharp time separation of ions in various charge states or with different $m/q$ (see Fig.~\ref{fig:chargebreeding}) and allows for the selection of ions with a certain ($m/q$)-ratio using a Bradbury-Nielsen ion gate \cite{Brunner2012}.
A distribution of charge-bred $^{85}$Rb-ions extracted from the EBIT is shown in Fig.~\ref{fig:chargebreeding}. 
In this case a charge-breeding time of 197~ms was used.
The HCI with the charge state taken to be the one with the greatest ratio of ion of interest to background are sent to the MPET to determine the ion's mass using the time-of-flight ion cyclotron resonance (TOF-ICR) method \cite{Graff1980, Konig1995}.

In the MPET ions are trapped radially via a strong homogeneous magnetic field $B$ and axially via a weak harmonic electric field. 
The mass of the ion of interest $M$ in the charge state $q$ is determined from the ion's cyclotron frequency
\begin{equation}
	\nu_{\rm c} = \frac {q \cdot B}{2 \pi \cdot M} .
	\label{eq:nuc}
\end{equation}
\noindent A direct measurement of $\nu_{\rm c}$ is done by applying a quadrupolar radio-frequency field $\nu_{\rm rf}$ which converts the radial magnetron motion into the reduced cyclotron motion (for details on the application of the technique at TITAN see \cite{Brodeur2012}).
When $\nu_{\rm rf}$ is equal to $\nu_{\rm c}$ the ions gain radial energy, which is converted into longitudinal energy when extracted from the trap through the magnetic-field gradient and subsequently detected with a microchannel plate detector. The resonant ions therefore have a shorter time of flight. A frequency range around the expected $\nu_{\rm c}$ is scanned to obtain a resonance (Fig.~\ref{fig:resonance94Rb} and \ref{fig:resonance98Rb}).

The statistical uncertainty on a mass measurement is given by
\begin{equation}
	\frac{\delta m}{m} = \mathcal{F} \cdot \frac{m}{q \cdot B \cdot T_{\rm RF} \cdot \sqrt{N_{\rm ion}}},
	\label{eq:precision_mass}
\end{equation}
\noindent where $m$ is the atomic mass, $T_{\rm RF}$ is the excitation time, $N_{\rm ion}$ is the number of ions sampled in the measurement \cite{Bollen2001}, and $\mathcal{F}$ is a trap specific parameter which is close to 1 for TITAN. The excitation time $T_{\rm RF}$ is limited by the half-life of the radioactive ions.
TITAN can boost the precision of the mass measurement, which scales linearly with $q$, by increasing the charge state of the ions.
Additionally, the increase in resolving power due to charge breeding allows the resolution of low-lying nuclear isomers in Penning traps \cite{Gallant2012}. 
Alternatively, 
the integrated measurement time can be reduced due to the use of HCI as shown by this work. For radioactive ions with low yields, high charge states allow for a decrease in the total number of measured ions while still maintaining high precision. 
In order to assess the benefits of charge breeding, additional aspects have to be considered. 
The charge breeding process is accompanied by loss mechanisms resulting from ion transport, capture, storage, and extraction. Furthermore, the ions are distributed over several charge states, the energy spread is increased, and the additional time delay increases decay losses. The probability of charge exchange with residual gas during the excitation in the Penning trap becomes higher. However, for short lived nuclei this excitation time has to be kept short. As can be seen in Fig.~\ref{fig:resonance94Rb} the quality of the resonance is not significantly diminished at $T_{\rm RF}=97$~ms.
As demonstrated at SMILETRAP \cite{Bergstrom2002} with highly charged ions of stable nuclei, trapping times of up to several seconds are possible without significant charge exchange losses.

\section{Data analysis}

\begin{table*}
	\caption{\label{tab:ME} Frequency ratios of $^{94,97,98}$Rb$^{15+}$ and $^{94,97-99}$Sr$^{15+}$ isotopes relative to $^{85}$Rb$^{13+}$ as well as mass excesses. Uncertainties are expressed in parenthesis. The first error on the frequency ratio represents the statistical error multiplied by the reduced chi square of the fit of the line shape, count-class analysis as well as time correlations. The second and third errors represent systematics related to $\Delta (m/q)$-dependent shifts and ambiguities in the choice of time-of-flight range. The fourth error in square brackets represents the quadrature sum of all the errors. For the mass excesses the combined uncertainty is shown. In the last two columns the mass excess values from AME03 \cite{Audi2003b} and JYFLTRAP (Rb \cite{Rahaman2007}, Sr \cite{Hager2006}) are listed.}
\begin{tabular*}{1.0\textwidth}{@{\extracolsep{\fill}} l l l l l l }
	\hline
	\hline
	Isotope 			& $T_{1/2}$	& $\overline{R} = \nu_{\rm c,ref} / \nu_{\rm c}$ 			& $ME_{\rm TITAN}$ (keV) 		& $ME_{\rm AME03}$ (keV)	& $ME_{\rm JYFLTRAP}$ (keV) \\ 
	\hline 
	\hline
	$^{94}$Rb$^{15+}$  	& 2.702 s	& 0.958672311(22)(16)(3)[27]	& -68562.6(2.4) 	& -68553.4(8.4)	& -68564(5) \\
	$^{97}$Rb$^{15+}$ 	& 169.9 ms	& 0.989404952(17)(16)(0)[23]	&  -58519.2(2.1)	& -58356.3(30.5)	& -58519(6) \\
	$^{98}$Rb$^{15+}$ 	& 114 ms	& 0.999658513(31)(16)(14)[38]	& -54318.4(3.4) 	& -54221.6(50.2) 	& \\
	\hline
	$^{94}$Sr$^{15+}$ 	& 75.3 s	& 0.958559623(10)(16)(0)[19]	& -78845.8(1.7)	& -78840.4(7.2) 	& \\
	$^{97}$Sr$^{15+}$ 	& 429 ms	& 0.989294688(37)(16)(2)[40]	& -68581.2(3.6)	& -68788.1(19.2) 	& -68587(10) \\
	$^{98}$Sr$^{15+}$ 	& 653 ms	& 0.999525849(41)(16)(5)[44]	& -66424.5(4.0)	& -66645.7(26.3) 	& -66431(10) \\
	$^{99}$Sr$^{15+}$ 	& 269 ms	& 1.009776308(42)(16)(3)[45]	& -62506.8(4.1)	& -62185.7(80.0) 	& -62524(7) \\
	\hline
	\hline
\end{tabular*}
\end{table*}

The main observable in Penning-trap mass spectrometry is the cyclotron frequency $\nu_{\rm c}$ (see Eq.~\ref{eq:nuc}). The charge state $q$ is derived from a time-of-flight spectrum, but the magnetic field needs to be measured.
From a fit of the theoretical line shape \cite{Konig1995} to the resonance data (see Fig.~\ref{fig:resonance94Rb} and \ref{fig:resonance98Rb}), the mass can be extracted from Eq.~\ref{eq:m} if $q$ and $B$ are known. 
To minimize systematic effects (discussed below) and to calibrate the magnetic field, a reference measurement of an ion with a well-known mass and $q$ is performed before and after the actual measurement.
The primary experimental result is the ratio of the cyclotron frequency of the ion of interest to that of the reference ion $\nu_{\rm c,ref}$, thus the ratio of the masses,
\begin{equation}
	R = \frac{\nu_{\rm c,ref}} {\nu_{\rm c}} = \frac{q_{\rm ref}\cdot M} {q \cdot M_{\rm ref}}.
	\label{eq:R}
\end{equation}

\noindent The atomic mass $m$ is given by:
\begin{equation}
	m = \frac{q}{q_{\rm ref}} \cdot \overline{R} \cdot (m_{\rm ref} - q_{\rm ref} \cdot m_{\rm e} + B_{\rm e,ref}) + q\cdot m_{\rm e} - B_{\rm e},
	\label{eq:m}
\end{equation}
\noindent where $\overline{R}$ is the average of all measured frequency ratios, $B_{\rm e}$ is the total binding energy of the removed electrons (also known as neutralization energy) of the ion of interest, and the index ``$\rm ref$'' refers to the reference ion.

The value for $\nu_{\rm c,ref}$ at the time of the measurement of $\nu_{\rm c}$ is obtained by a linear interpolation of the two reference measurements that enclose the measurement of the ion of interest. 
There is a correlation introduced between adjacent frequency ratio measurements due to shared references. For the data analysis we take into account a full covariance matrix between all the ratios \cite{Valassi2003}.

In Tab.~\ref{tab:ME} the frequency ratios of $^{94,97,98}$Rb$^{15+}$ and $^{94,97-99}$Sr$^{15+}$ isotopes relative to $^{85}$Rb$^{13+}$ as well as mass excesses of the atomic mass are presented. 
Accuracy limitations, reflected in systematic uncertainties, such as the following are considered:

\begin{itemize}
\item A spatially-uniform magnetic field is required. We minimize the effect of instabilities in the magnetic field by using a frequency ratio where several systematics cancel out. As stated in \cite{Brodeur2012} the uncertainty for magnetic field instabilities is $\Delta R/ R \ll$0.2~ppb/ hour between adjacent reference measurements. In this work the time between two reference measurements was kept to less than one hour. 

\item The uncertainty in the mass of the reference ion can be considered negligible, since we use $^{85}$Rb which is known with a mass uncertainty of 11~eV \cite{Mount2010}, equivalent to 0.1~ppb.

\item The total electron binding energies were taken from \cite{Rodrigues2004}. There the total atomic binding energy (binding energy of all remaining electrons) has been calculated using a Dirac-Fock approximation and values are tabulated for lithium- to dubnium-like systems with $Z=3..118$. Uncertainties for the total electron binding energies of Rb$^{13+, 15+}$ and Sr$^{15+}$ are conservatively estimated to be below 20~eV, corresponding to 0.2~ppb \cite{Indelicato2011}.

\item Although we aim for single-ion injection, the time-of-flight spectra of the ions detected on the microchannel plate detector after the Penning trap show a multitude of ($m/q$)-states. These are unresolved charge states due to charge exchange with residual background gas in the MPET. 
To minimize the possibility of charge exchange a low pressure is favored. The pressure in the MPET vacuum section was measured to be $\approx5\cdot10^{-11}$~mbar. 
The third parenthesis in Tab.~\ref{tab:ME} represents the uncertainty associated with gating on the charge state of interest. 
By varying the time-of-flight range in the analysis considering a worst-case scenario, the uncertainty is estimated to be a few ppb.

\item If more than one ion is stored in the trap we account for shifts due to ion-ion interactions using a count-class analysis \cite{Kellerbauer2003}. 
This uncertainty is convoluted with the statistical uncertainty (see first parenthesis in Tab.~\ref{tab:ME}) and varies from 0 to 3~ppb depending on the isotope. 

\item Accuracy checks for HCI needed in this work have been performed by measuring the cyclotron frequency of $^{85}$Rb$^{11+}$ vs. $^{85}$Rb$^{13+}$ vs. $^{85}$Rb$^{15+}$. Different settings (e.g. Lorentz steerer \cite{Ringle2007}, extraction optics, etc.) as well as different timings (e.g. duty cycle, capture timings, etc.) were covered. 
A shift in frequency cannot be excluded due to possible trap misalignment, ion-ion interactions due to the increase of charge states, and relativistic mass increase. The mass shift due to the increase of charge states scales with 
the difference in the ratio of $m/q$ of ion of interest to reference, $\Delta (m/q)$, the relativistic effect in this measurement setup with $\Delta (q/m)$. 
The systematic uncertainty for these checks is conservatively estimated to be $<15.5$~ppb (absolute) and is indicated in Tab.~\ref{tab:ME} in the second parenthesis. 
Therefore a small $\Delta$ in the ratio of $m/q$ or $q/m$ of ion of interest to reference is essential. We used $^{85}$Rb$^{13+}$ with $(m/q)=6.5$ as the reference ion to minimize $\Delta (m/q)$ effects.

\item Systematic shifts due to isobaric contamination needs to be minimized.
The Sr beam was free of contaminations whereas in the case of all Rb mass measurements, not all contaminating ions could be removed using ISAC's mass separator. We used a dipole radio-frequency excitation \cite{Blaum2006} preceding the quadrupole frequency scan to remove all contaminations. These were identified to be the Sr isobars. The application of a dipole (reduced cyclotron) frequency of the contaminant in the radial plane results in the excitation of their cyclotron radii. 
During the extraction of the ions the excited ions will not clear the aperture. This is a process referred to as dipole cleaning.
\end{itemize}

\section{Results}

The results of the mass measurements of  $^{94,97,98}$Rb and $^{94,97-99}$Sr performed with TITAN are summarized in Tab.~\ref{tab:ME} and discussed in subsections~\ref{94Rb}-\ref{94to99Sr}. 
The absolute uncertainty of all investigated isotopes is below 4~keV, including the new direct mass measurement of $^{98}$Rb. In Tab.~\ref{tab:ME} column $ME_{\rm TITAN}$ lists the measured TITAN mass excess with the different individual uncertainties, as well as the combined uncertainty. The next two columns show results found in the literature for these isotopes. The visual comparison between the mass excess by TITAN and the previous measurements is shown in Fig.~\ref{fig:results}. A global mass evaluation as outlined in \cite{Audi2003b} was performed and the evaluated values for the ions of interest in this work can be seen in Tab.~\ref{tab:influences} and later in the text.

\subsection{\label{94Rb}$^{94}$Rb}
\begin{figure}
    \begin{center}
	\includegraphics[width=8.6cm]{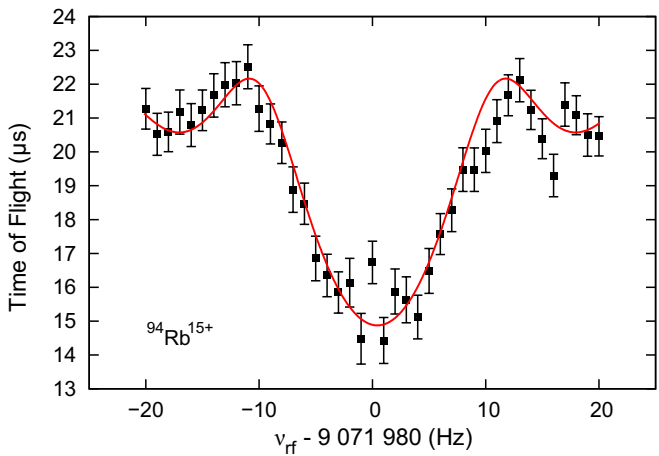} 
    \end{center}
    \caption{\label{fig:resonance94Rb}(color online) $^{94}$Rb$^{15+}$ cyclotron resonance taken with 80~ms charge-breeding time, 20~ms dipole cleaning, and 77~ms excitation time in MPET. The solid line is a fit of the theoretical line shape \cite{Konig1995} to the data.}
\end{figure}

\begin{figure*}
    \begin{center}
	\includegraphics[width=8.6cm]{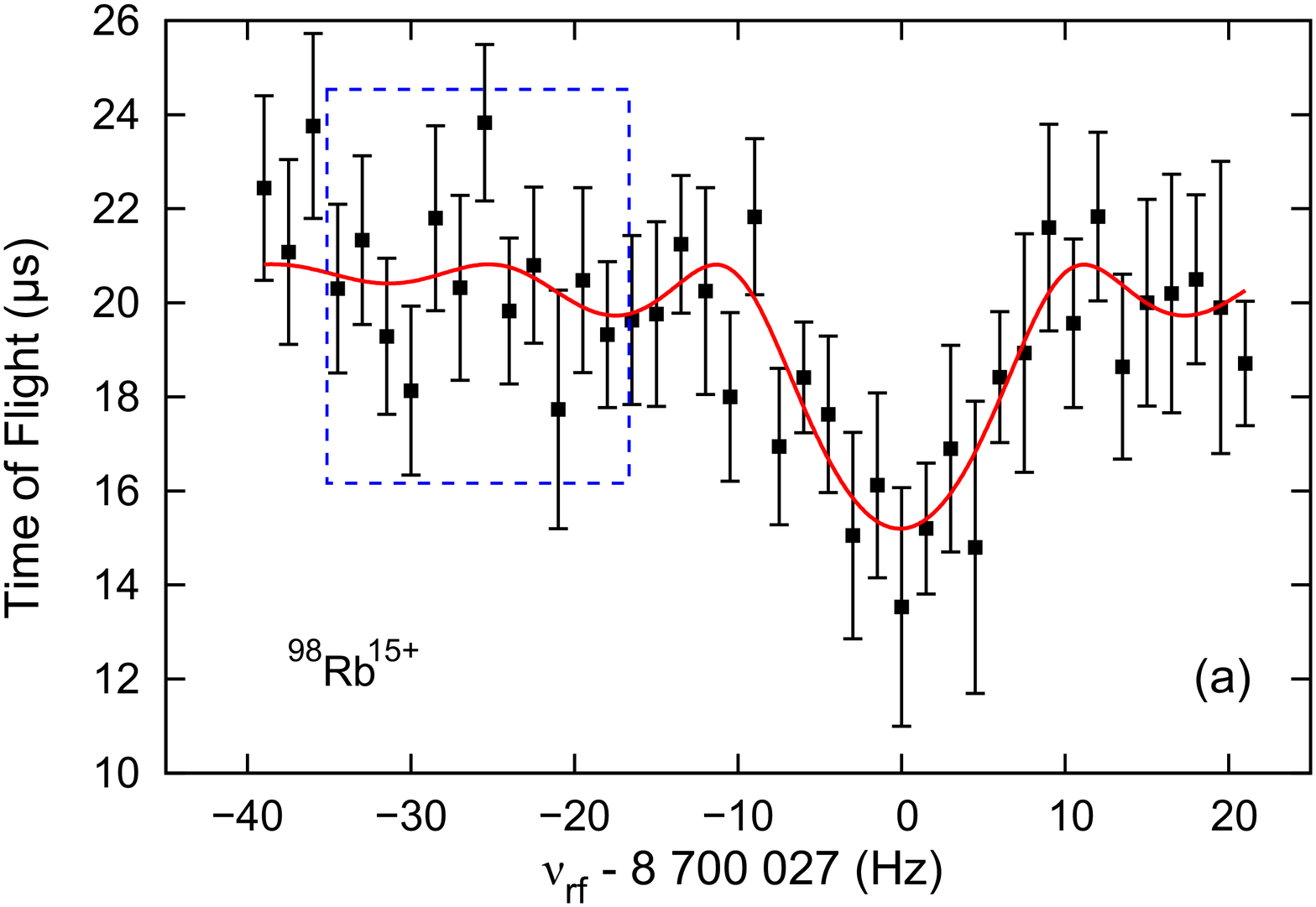} 
	\quad
	\includegraphics[width=8.6cm]{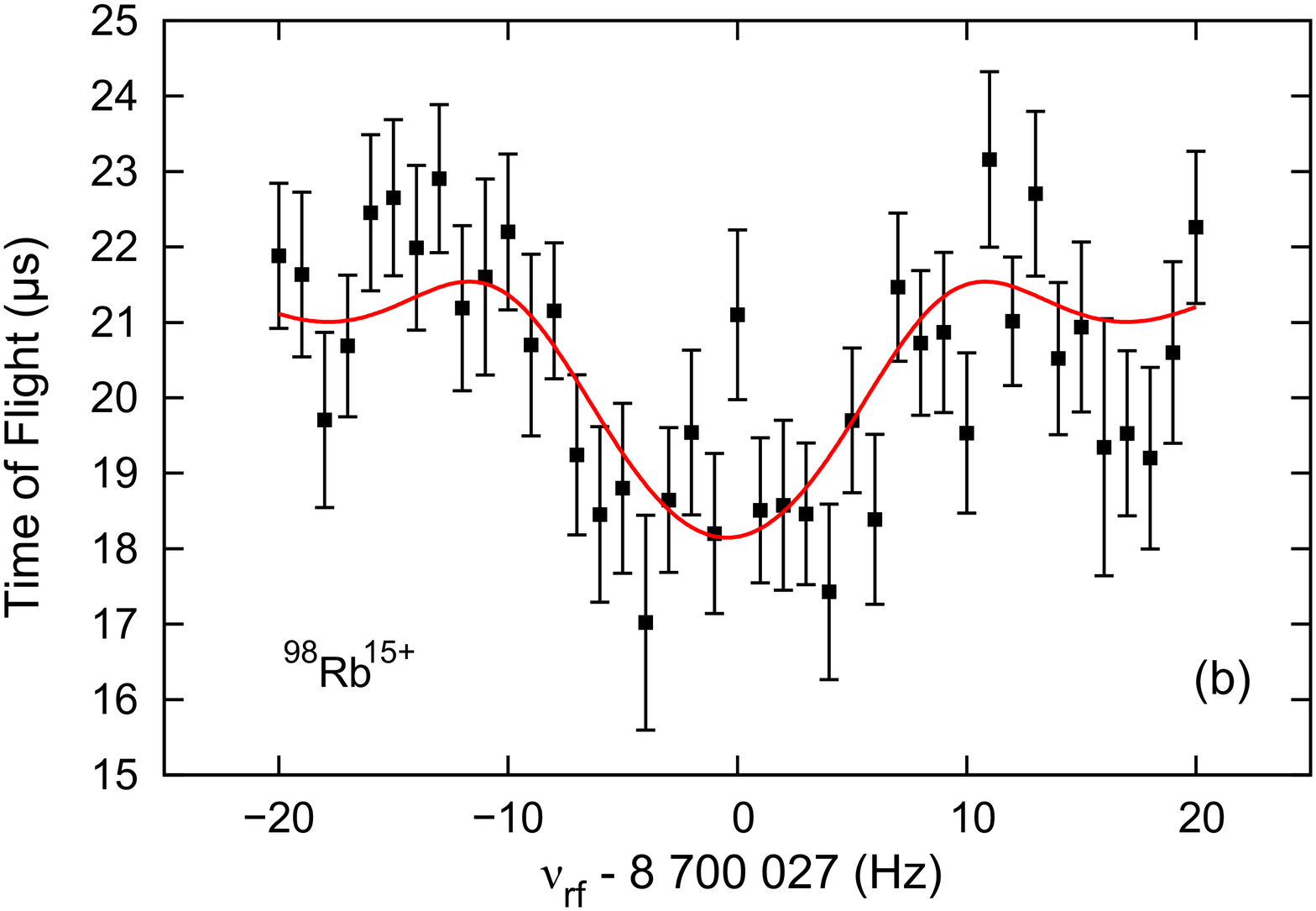} 
    \end{center}
    \caption{\label{fig:resonance98Rb}(color online) The cyclotron resonance for $^{98}$Rb$^{15+}$ was taken with 80~ms charge-breeding time, 20~ms dipole cleaning to eliminate $^{98}$Sr$^{15+}$, and 77~ms excitation time in MPET. The solid line is a fit of the theoretical line shape \cite{Konig1995} to the data. The left graph shows the resonance off center to include the range of the proposed isomer $^{98}$Rb$^m$ in frequency space indicated by a dashed blue square. The graph on the right shows another measurement with a centered resonance. See Tab.~\ref{tab:ME} for results.}
\end{figure*}

The measurement campaign began with $^{94}$Rb ($T_{1/2}=$2.702(5)~s) which is relatively well known according to AME03 \cite{Audi2003b}, $\delta m=8.4$~keV. A cyclotron resonance of $^{94}$Rb$^{15+}$  is shown in Fig.~\ref{fig:resonance94Rb}.
With our mass excess of -68562.6(2.4)~keV we find excellent agreement not only with AME03 (see Tab.~\ref{tab:ME}) but also with the Penning-trap mass measurements at ISOLDE \cite{Raimbault-Hartmann2002} and JYFLTRAP \cite{Rahaman2007}.

\begin{figure}
    \begin{center}
	\includegraphics[width=8.6cm]{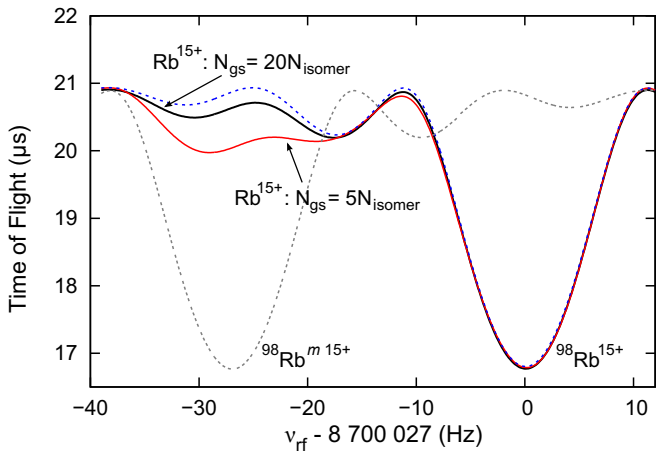} 
    \end{center}
    \caption{\label{fig:isomer}(color online) This calculation displays the theoretical line shape of the isomer  $^{98}$Rb$^{m\,15+}$ in dashed gray and the ground state $^{98}$Rb$^{15+}$ in dashed blue if only one of them is present. If both are present the solid lines indicate the line shapes for the resonance. If the yield of the isomer is a factor of 20~(5) less than for the ground state, the resonance line shape is displayed with a solid black (red) line.}
\end{figure}

\subsection{\label{97Rb}$^{97}$ Rb}
Our measured mass excess for $^{97}$Rb ($T_{1/2}=$169.9~ms) of -58519.2(2.1)~keV lies within the uncertainty of the JYFLTRAP value \cite{Rahaman2007}, but it differs by 163~keV (5.3$\sigma$) from the value adopted by AME03 \cite{Audi2003b}. We confirm the JYFLTRAP value with an improved precision. 
The previous measurements considered in AME03 were based on $\beta$ end-point energies from $^{97}$Rb($\beta ^{-}$)$^{97}$Sr. The adjusted $Q$-value, including the TITAN input, is now 10063(4)~keV, compared to the other measurements: 10020(50)~keV \cite{Decker1980} (0.9$\sigma$ deviation), 10450(30)~keV \cite{Blonnigen1984} (12.9$\sigma$ deviation), 10440(60)~keV \cite{Graefenstedt1987} (6.3$\sigma$ deviation), and 10462(40)~keV \cite{Przewloka1992} (10.0$\sigma$ deviation). 
The TITAN mass value validates the first measurement \cite{Decker1980} and greatly improves the precision of the $Q$-value and mass.

\begin{figure*}
    \begin{center}
	\includegraphics[width=8.6cm]{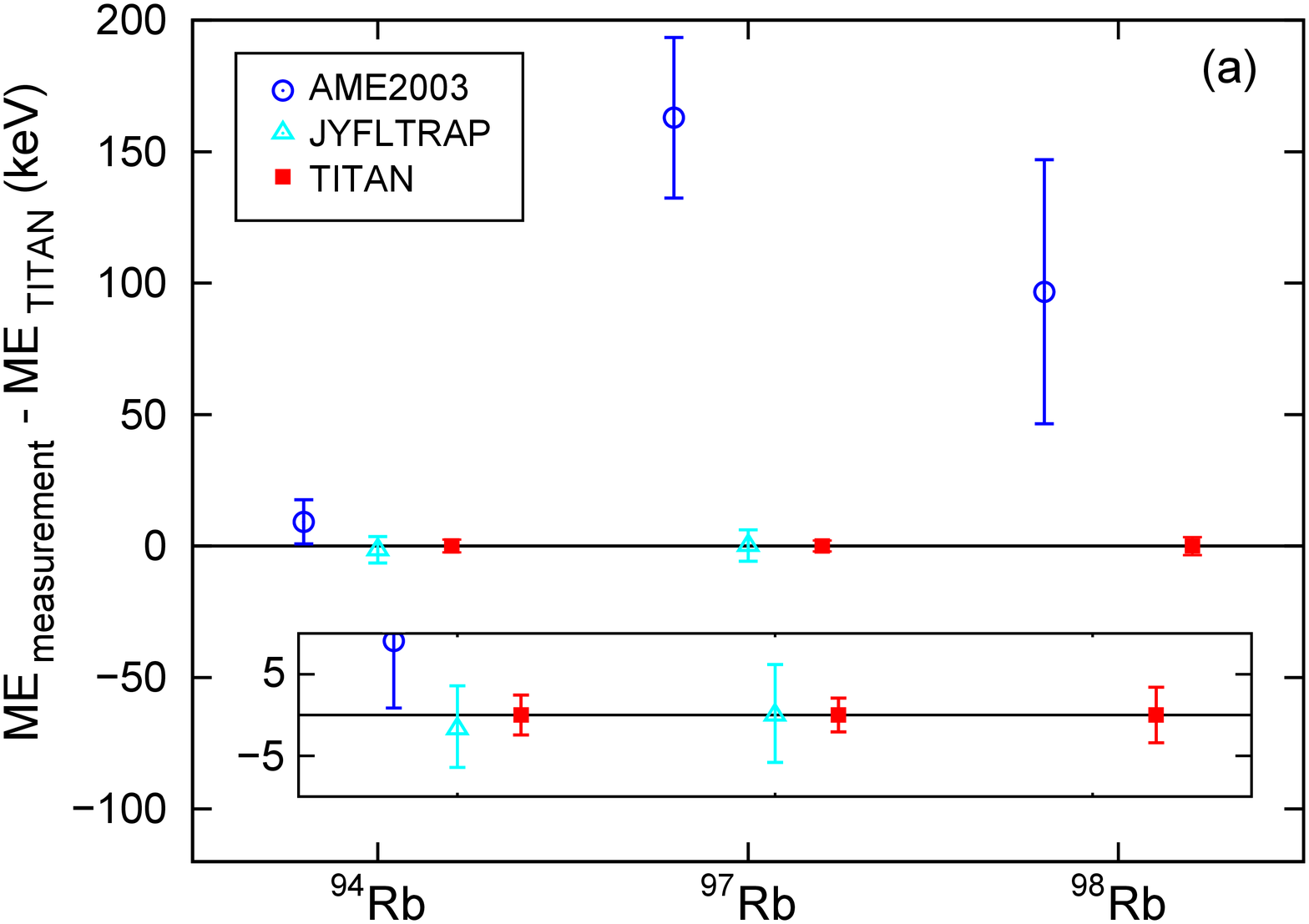} \quad 
	\includegraphics[width=8.6cm]{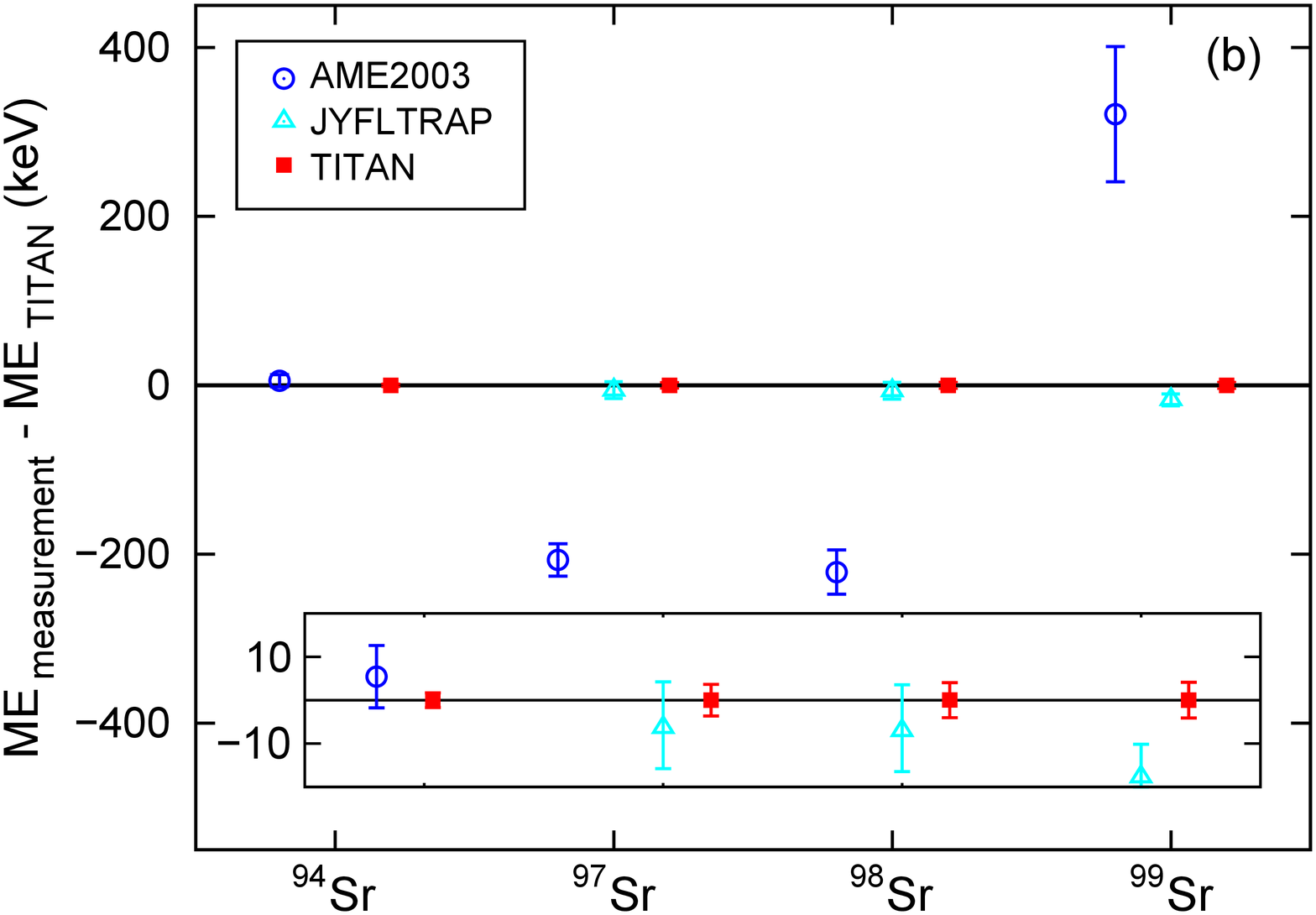} 
    \end{center}
    \caption{\label{fig:results}(color online) Comparison of the mass excesses determined in the present work (red squares), with JYFLTRAP \cite{Hager2006, Rahaman2007} (turquoise triangles) and in AME03 \cite{Audi2003b} (blue circles) for (a) Rb isotopes and (b) Sr isotopes. The inset displays an enlarged view.}
\end{figure*}

\subsection{\label{98Rb}$^{98}$Rb (ground and isomeric state)}
This work presents the first Penning-trap mass measurement of $^{98}$Rb ($T_{1/2}=$114(5)~ms). Our measured mass excess is -54318.4(3.4)~keV and differs from the adopted value in AME03 \cite{Audi2003b} of -54221.6(50.2)~keV by 97~keV or 1.9$\sigma$ deviation. The previous mass excess was determined from the end-point energy of the $\beta$ spectrum of $^{98}$Rb($\beta^{-}$)$^{98}$Sr \cite{Graefenstedt1987, Przewloka1992} with 80.4\%  weight and from the triplet measurement of ($^{97}$Rb, $^{98}$Rb, $^{95}$Rb) \cite{Audi1986} with 19.6\% weight.

A low-lying isomeric state in $^{98}$Rb is proposed at 286(128)~keV \cite{Audi2011} with a half-life of 96(3)~ms \cite{NationalNuclearDataCenter2012}. 
The adopted value in AME03 takes into account not only $Q$-values from $\beta$ end-point energies as mentioned for $^{98}$Rb, but also the mass-triplet measurement of reference \cite{Audi1986} for the ground state combined with $^{98}$Rb$^m$($\beta^{-}$)$^{98}$Sr \cite{Graefenstedt1987}. 
In an effort to confirm the energy of the isomer, we scanned a range of 630~keV (see Fig.~\ref{fig:resonance98Rb}a), in which we expected to observe the isomer. If the isomer were present, two dips would have been visible in the resonance curve. Only one dip, consistent with the ground state, was observed; however, the absence of a second resonance does not exclude the possibility of an isomer. The strength of the resonances depends on the ratio of the population of the isomeric and ground states. Yield measurements at ISAC \cite{Dombsky2012} indicated the yield of the ground state to be 20 times larger than the isomeric state. If this ratio were observed with MPET, the signal is expected to be the black, solid curve in Fig.~\ref{fig:isomer}, from which the isomer cannot be detected. If, however, there were only five times more ions in the ground state than in the isomeric state, the isomeric state would be detectable (red, solid curve). To guide the eye, a resonance is drawn as if each state alone were trapped and measured with a dashed curve (gray for the isomer and blue for the ground state).
To further confirm the energy of the isomer a higher charge state, preferably an isoelectronic series of Ar or Ne corresponding to $q=19+$ and $q=27+$ respectively, could be used for the mass measurement to strongly enhance the resolving power. 
This would allow the implementation of dipole cleaning of the ground state to enhance the resonance of the isomer.

\subsection{\label{94to99Sr}$^{94,97,98,99}$Sr}
For all the measured Sr masses the uncertainties were reduced.
The mass excess of $^{94}$Sr is known to 7~keV accuracy from measurements at ISOLDE \cite{Audi2003b}. Our measurement agrees and improves the accuracy by a factor of 4 to 1.7~keV as presented in Tab.~\ref{tab:ME} and Fig.~\ref{fig:results}. 
For $^{97}$Sr our measured mass differs by 207~keV and 10.8$\sigma$ deviation to AME03 \cite{Audi2003b}, but it confirms the mass measurement from JYFLTRAP \cite{Hager2006}. 
In AME03 $^{97}$Sr is mainly determined from $^{97}$Sr($\beta ^{-}$)$^{97}$Y. The adjusted $Q$-value, including the TITAN input, is now 7545(8)~keV. The previously measured $Q$-value was underestimated: 
7452(40)~keV \cite{Blonnigen1984} (2.3$\sigma$ deviation), 7420(80)~keV \cite{Graefenstedt1987} (1.6$\sigma$ deviation), and 7480(18)~keV \cite{Groß1992} (3.6$\sigma$ deviation).
The mass from TITAN is significantly more precise than previous Penning-trap mass measurements by a factor of 3.

The scenario for $^{98}$Sr is similar where our value with -66424.5(4.0)~keV is in agreement within the error of JYFLTRAP \cite{Hager2006} and more precise, but disagrees with AME03 \cite{Audi2003b} by 221~keV (8.4$\sigma$ deviation).
AME03 adopted its value from $\beta$ end-point energy experiments, with 95.5\% from $^{98}$Sr($\beta ^{-}$)$^{98}$Y \cite{Blonnigen1984} and 4.5\% from $^{98}$Rb($\beta ^{-}$)$^{98}$Sr \cite{Graefenstedt1987, Przewloka1992}. 

In the case of $^{99}$Sr where the mass potentially plays an important role for the $r$-process and further mass extrapolations, we obtain a mass excess of -62506.8(4.1)~keV. This value agrees within 2.5$\sigma$ with the JYFLTRAP measurement \cite{Hager2006}, but it disagrees by 321~keV (4.0$\sigma$) with the mass evaluation \cite{Audi2003b}. This is the only case where we differ from the JYFLTRAP measurements. 
Our mass excess gravitates slightly away from their measurement towards AME03 \cite{Audi2003b} where the adopted mass stems from $\beta$ end-points from $^{99}$Sr($\beta ^{-}$)$^{99}$Y \cite{Iafigliola1984}, with 91\% weight and $^{99}$Rb($\beta ^{-}$)$^{99}$Sr \cite{Iafigliola1984}, with 9\% weight.

\begin{table*}
	\caption{\label{tab:influences}Most important contributing data to and their influences on its mass as it will appear in the next mass evaluation \cite{Audi2011} following AME03 \cite{Audi2003b} and extended by the TITAN masses from this work. Influences are given as a percentage. Experimental techniques displayed such as ($^{94}$Rb vs. $^{85}$Rb) indicate Penning-trap mass spectrometry and ($^{97}$Rb$(\beta ^{-})$$^{97}$Sr) $\beta$ end-point energy experiments.}
	\begin{tabular*}{1.0\textwidth}{@{\extracolsep{\fill}} l |l l |l l |l }
	\hline
	\hline
	Nucleus \quad& Influence	& TITAN				& Influence		& Others experiments	\quad					& Evaluated $ME$ (keV)	 \\ 
	\hline 
	\hline
	$^{94}$Rb  	& 70.2\%	& $^{94}$Rb vs. $^{85}$Rb \quad 	& 29.6\%	& $^{94}$Rb vs. $^{88}$Rb  \cite{Rahaman2007} 		& -68562.3(2.0)	\\
	$^{97}$Rb  	& 87.0\%	& $^{97}$Rb vs. $^{85}$Rb  	& 12.9\% 		& $^{97}$Rb vs. $^{88}$Rb  \cite{Rahaman2007}		& -58518.5(1.9)	\\
	$^{98}$Rb  	& 100\%	& $^{98}$Rb vs. $^{85}$Rb  	& 			& 									& -54317.7(3.4)	\\
	\hline
	$^{94}$Sr 	& 98.4\%	& $^{94}$Sr vs. $^{85}$Rb 	& 1.6\%		& $^{94}$Sr$(\beta ^{-})$$^{94}$Y  \cite{Decker1980} 	& -78845.1(1.7)	\\
	$^{97}$Sr 	& 87.4\%	& $^{97}$Sr vs. $^{85}$Rb  	& 12.6\% 		& $^{97}$Sr vs. $^{97}$Zr  \cite{Hager2006} 		& -68581.8(3.4)	\\
	$^{98}$Sr 	& 85.2\% 	& $^{98}$Sr vs. $^{85}$Rb  	& 14.8\%		& $^{98}$Sr vs. $^{97}$Zr  \cite{Hager2006}		& -66425.6(3.7)	\\
	$^{99}$Sr 	& 75.9\%	& $^{99}$Sr vs. $^{85}$Rb  	& 24.0\%		& $^{99}$Sr vs. $^{99}$Zr  \cite{Hager2006} 		& -62511.7(3.6)	\\
	\hline
	\hline
\end{tabular*}
\end{table*}

\begin{figure*}
    \begin{center}
	\includegraphics[width=8.6cm]{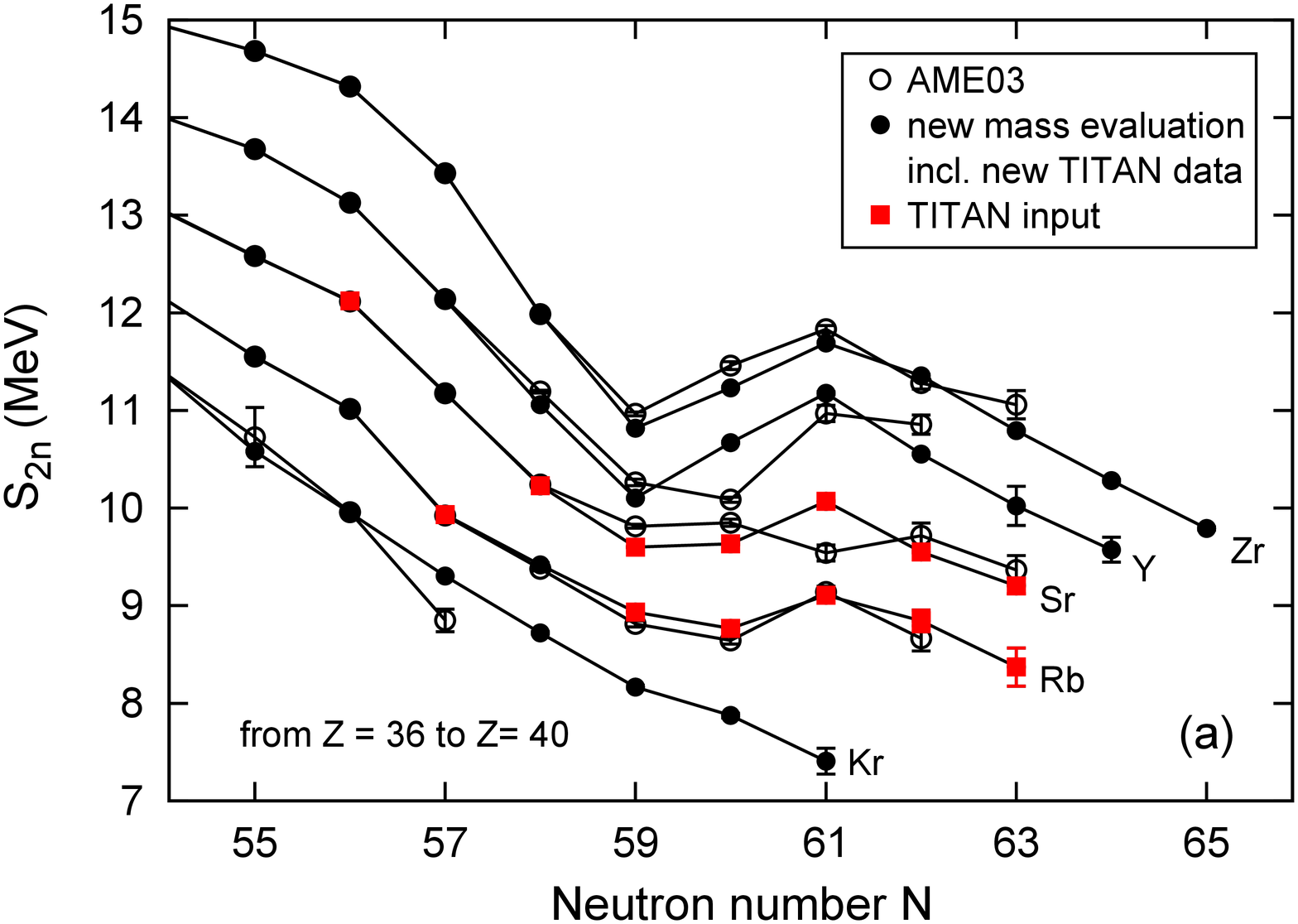} \quad 
	\includegraphics[width=8.6cm]{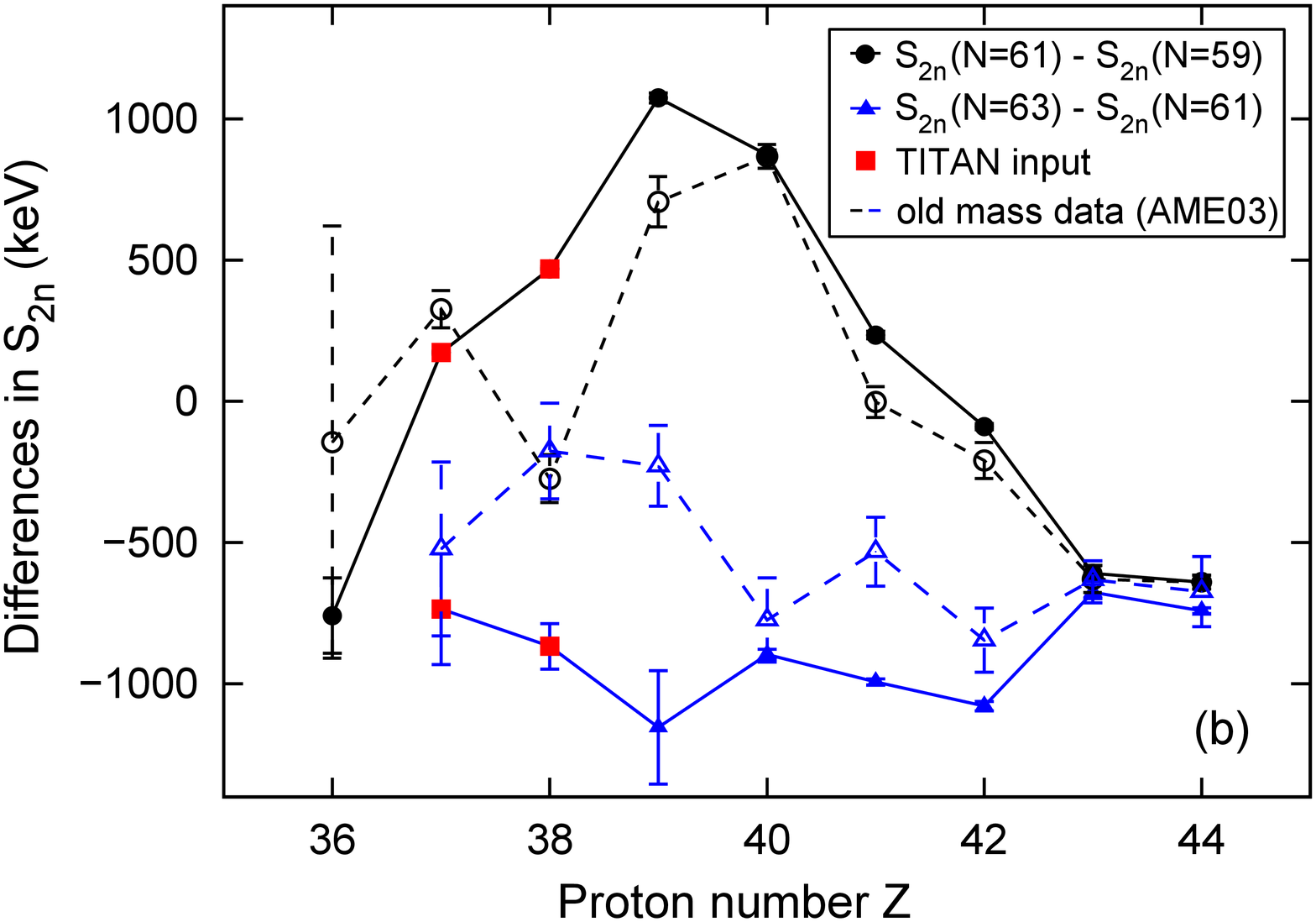}	
    \end{center}
    \caption{\label{fig:S2n_Y}(color online) (a): Two-neutron separation energies (S$_{2n}$) for $ Z=36-40$ (Kr to Zr) versus neutron number. In open black circles data from AME03 \cite{Audi2003b} is presented. For comparison in filled black circles results from recent mass spectrometry experiments from JYFLTRAP \cite{Hager2006, Hager2007, Rahaman2007} and ISOLTRAP \cite{Naimi2010, Delahaye2006} are shown extended by the TITAN masses. The TITAN mass input itself is indicated by red squares. 
(b): Differences between the S$_{2n}$ of the isotones $N=61$, $N=59$ in black circles, $N=63$, $N=61$ in blue triangles, and previous AME03 \cite{Audi2003b} data in dashed lines versus the proton number. This observable illustrates the quantum nuclear shape transition between $N=61$ to $N=59$ with Kr and Tc, Ru being the lower and upper limits. In contrast the difference in S$_{2n}$ for $N=63$, $N=61$ displays the normal behavior again.}
\end{figure*}

\section{Discussion of the results}

The masses presented in this work impact nuclear structure and nuclear astrophysics studies.
The atomic mass evaluation adopted its final mass value from a weighting of mass values obtained from different experimental techniques. Therefore, links between nuclei exist, and our mass values influence other mass values as well. This is shown in a more systematic way in Tab.~\ref{tab:influences}. 
Listed here are the most important contributing data that are used to determine the mass of the nuclide as it will appear in the next mass evaluation \cite{Audi2011}. 
This upcoming mass evaluation follows AME03 \cite{Audi2003b}. It takes into account correlations between mass values due to relative mass measurements. To determine the evaluated mass value, one uses a least-squares method weighted according to the precision with which each piece of data is known. 
The measured Rb and Sr mass values not only influence their own isotope's evaluated value, but they also influence other mass values in the neutron-rich region and improve their precision due to links. 
In various isotopes e.g. $^{96}$Zr, $^{97}$Zr, $^{102}$Nb, and $^{104}$Nb the mass excess changes by one standard deviation. A global mass evaluation was carried out and the impact was investigated. However, a detailed investigation of  nuclei not measured in this work lies outside of the scope of this publication and can be seen in the new mass evaluation \cite{Audi2011}.

\subsection{Nuclear structure findings}
To assess the impact on nuclear structure, Fig.~\ref{fig:S2n_Y}~(a) shows the mass surface defined by the isotopic two-neutron separation energies S$_{2n}$.
It illustrates the deformation for $ Z=36-40$ (Kr to Zr) with data from AME03 in open black circles, recent mass spectrometry results from JYFLTRAP (for Rb, Sr, Y, Zr, Nb, and Mo masses \cite{Hager2006, Hager2007, Rahaman2007, Hakala2011}) and ISOLTRAP (for Kr masses \cite{Naimi2010, Delahaye2006}) in filled black circles, and the neutron-rich Rb, Sr data from TITAN (this work) in red squares. Extrapolated masses are disregarded.

The smooth trend of the S$_{2n}$ as seen for Kr (see Fig.~\ref{fig:S2n_Y}~(a, b) and \cite{Naimi2010}) is interrupted for other isotopic chains indicating a sudden change in deformation. 
The new data from this work agree with previous experiments showing an onset of large deformation for $A\approx100$ nuclei with $N\geq60 $. This can be strongly seen for Rb and Sr. 
Previous work indicates a rapidly changing behavior in nuclear structure in the region of $58\leq~N\leq61$. To visualize the deformation, the difference between the S$_{2n}$ of the isotones $N=61$, $N=59$ (black circles) and $N=63$, $N=61$ (blue triangles) versus the proton number is shown in Fig.~\ref{fig:S2n_Y} (b). This observable illustrates the so-called quantum nuclear shape transition \cite{Casten2009,Lhersonneau1994} between $N=61$ to $N=59$. A shape transition is clearly visible for Rb to Mo, whereas Kr presents the lower limit and Tc, Ru the upper limits, respectively. With the input from this work indicated in red squares, we obtain new data points extending to more neutron-rich isotopes. In contrast, the difference in S$_{2n}$ for $N=63$, $N=61$ displays the smooth behavior again. The slope of S$_{2n}$ forms its smooth trend again, which clearly indicates a strengthening of one nuclear shape.
For the Sr isotones from $N=61$ to $N=59$, previous data (AME03, dashed black lines in Fig.~\ref{fig:S2n_Y} (b)) showed no signature of unusual behavior, while our data and \cite{Hager2006} strongly display the shape transition in Sr.

\subsection{Astrophysical implications}

The masses measured in this work are relevant for a range of different types of $r$-process models. As an example we explore the astrophysical implications of our results with a parameterized, fully dynamic $r$-process model following Freiburghaus et al. \cite{Freiburghaus1999}. The model is inspired by the conditions that might be encountered in high entropy winds emerging from the nascent neutron star in a core collapse supernova explosion. 
As a starting point one assumes a fluid element that is heated to a very high temperature ($T\approx$9~GK) where the composition is essentially protons and neutrons, with the electron abundance $Y_e$ being set by weak interactions. The fluid element then undergoes a rapid expansion at constant velocity $v$, $Y_e$, and entropy $S$ ($S$ displayed in entropy per baryon in multiples of the Boltzmann constant). We choose similar model parameters to Hosmer et al. \cite{Hosmer2010}, i.e. $Y_e=0.45$ and a velocity $v=7500$~km/s. The model is coupled to a full reaction network with 5410 isotopes that includes all relevant charged particle, $\beta$ decay, and neutron capture rates that ensue. For unknown masses, mass extrapolations from \cite{Audi2003b} and calculated values from the finite-range droplet mass model (FRDM) \cite{Moller2012} were used. Masses enter exponentially in the calculation of ($\gamma$,n) photo disintegration rates from the forward (n,$\gamma$) rates via detailed balance. Calculations are carried out for a grid of entropies with the resulting isotopic abundances being added up with equal weight. Low entropies lead to low neutron-to-seed ratios and a weak $r$-process producing mainly lighter $r$-process isotopes, while higher entropies lead to more extended reaction paths all the way to the heaviest elements. An entropy range from 40--260 is sufficient to capture all entropies that contribute to the $r$-process. The attractive feature of this model is that it is inspired by the conditions one might encounter in high entropy winds from nascent neutron stars in core collapse supernovae, and that the solar system $r$-process abundance pattern can be reproduced reasonably well with just two free parameters -- $Y_e$ and $v$ \cite{Freiburghaus1999}. 

\begin{figure}
    \begin{center}
    	\includegraphics[width=8.6cm]{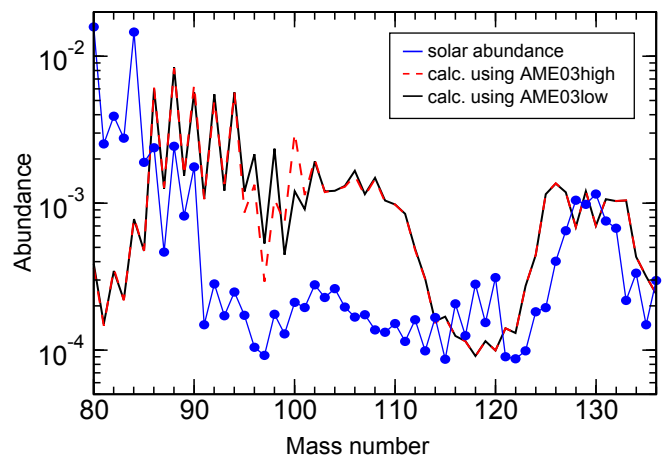} 
    \end{center}
           \caption{\label{fig:compare_nice}(color online) Calculated $r$-process abundances as a function of mass number summing all entropies for AME03high (dashed red) and AME03low (solid black) neutron separation energies for $^{97-99}$Rb and $^{97-100}$Sr. Also shown for comparison are the solar $r$-process residuals (filled blue circles) \cite{Travaglio2004}.}
\end{figure}

\begin{figure*}
    \begin{center}
    	\includegraphics[width=8.6cm]{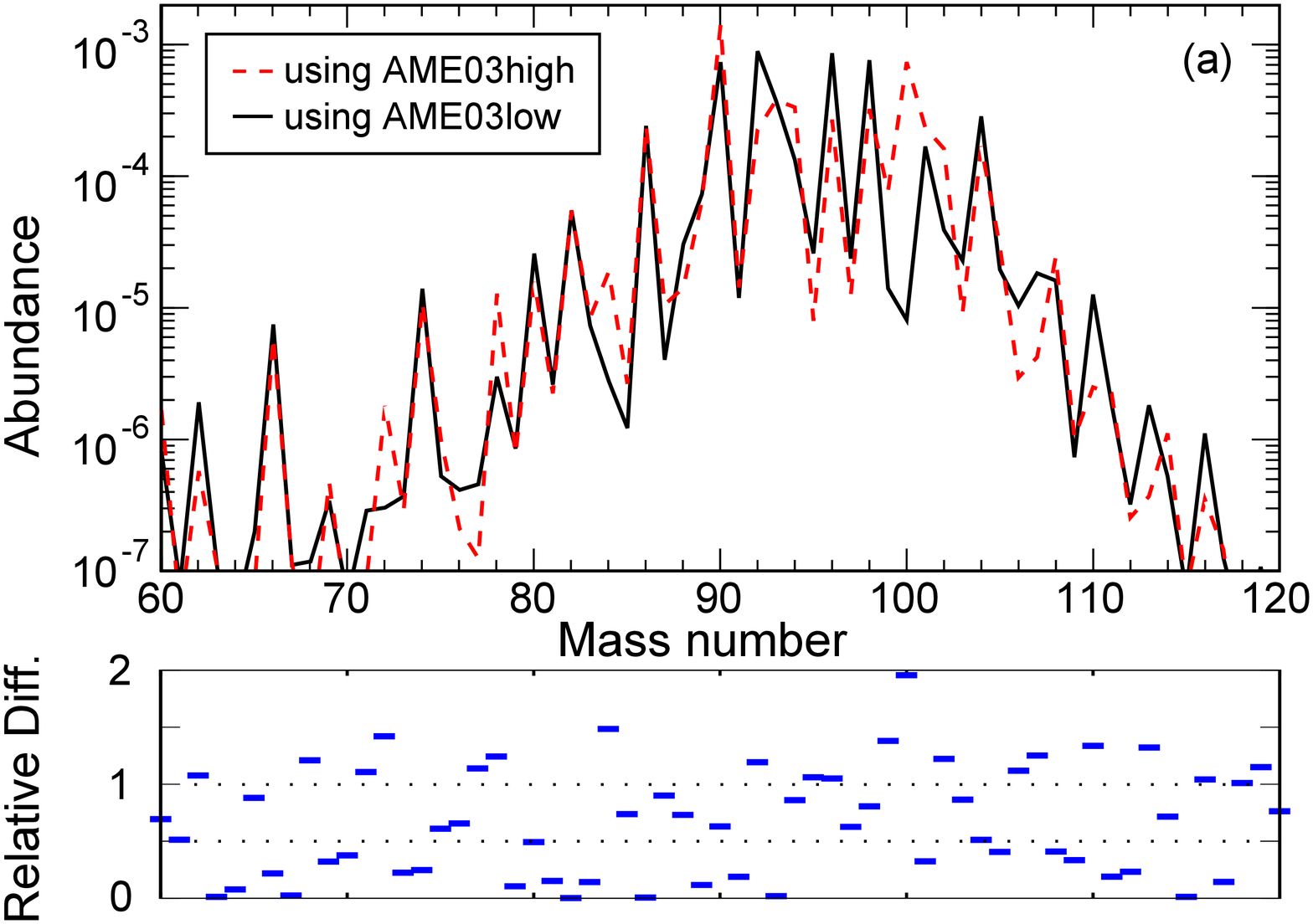} 
    	\includegraphics[width=8.6cm]{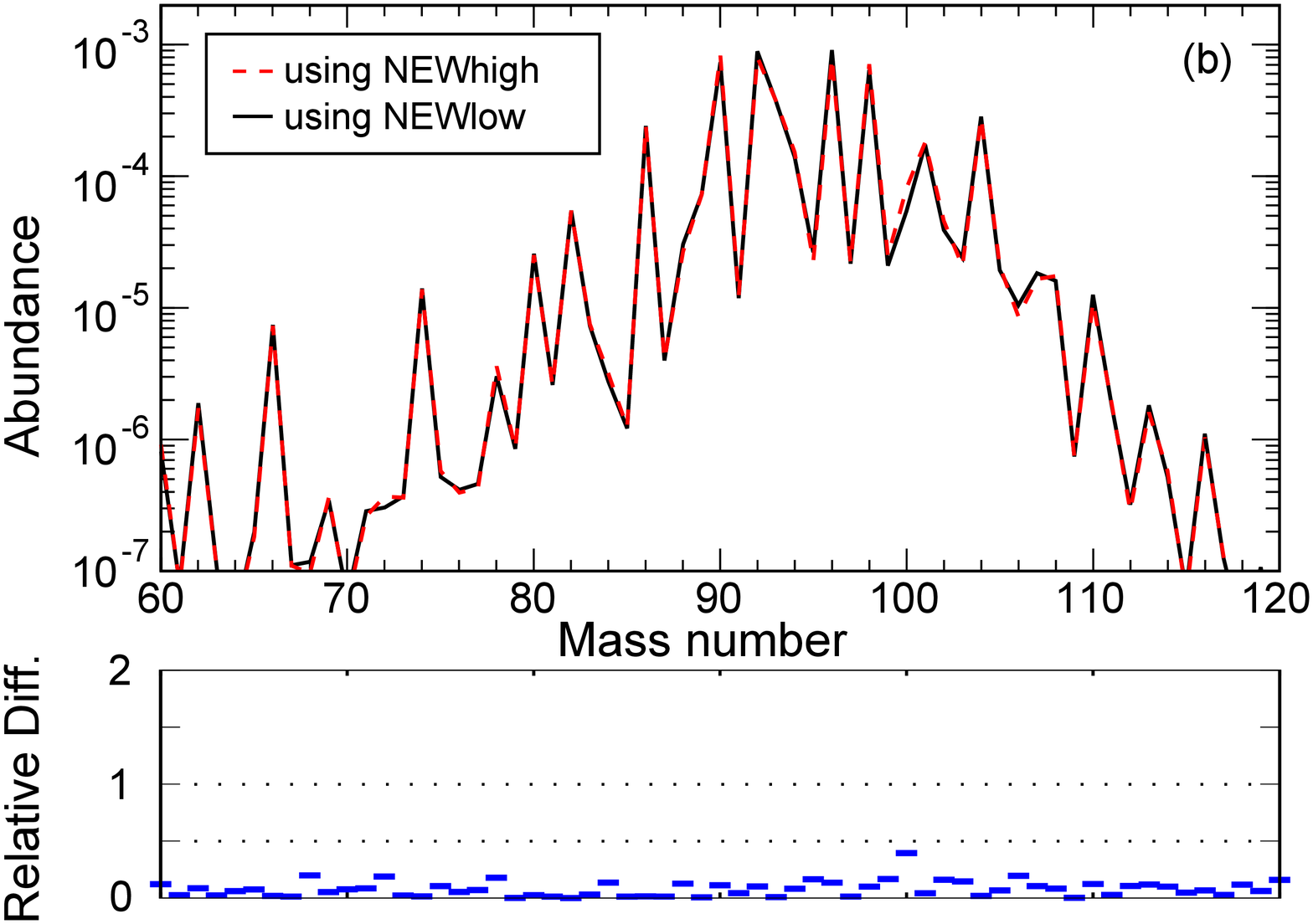} 
    \end{center}
           \caption{\label{fig:compare_90a_nice}\label{fig:compare_90b_nice}(color online) Calculated $r$-process abundances as a function of mass number for a single $S=100$ component for several data sets of neutron separation energies for $^{97-99}$Rb and $^{97-100}$Sr. 
           The left figure (a) shows results for variations in the neutron separation energies using AME03high (dashed red) and AME03low (solid black); The right figure (b) shows results using neutron separation energies from this work varied by the new experimental uncertainties with 3$\sigma$ (NEWhigh and NEWlow respectively). In addition in the lower panels, both figures display the relative difference between calculated abundances using the high and the low neutron separation energies
with $|\mathrm{high-low}|/[(\mathrm{high+low)}/2]$.}
\end{figure*}

To explore the relevance of the masses measured in this work for $r$-process simulations we performed two calculations where we varied the neutron separation energies of $^{96-99}$Rb and $^{96-100}$Sr either all up (AME03high) or all down (AME03low) by their AME03 \cite{Audi2003b} 3$\sigma$ errors. This results in variations of $\pm$40, $\pm$60, and $\pm$140~ keV for the Rb isotopes, respectively, and of $\pm$33, $\pm$33, $\pm$84, and $\pm$150~keV for the Sr isotopes, respectively. We choose a 3$\sigma$ variation as such deviations are not uncommon for non-Penning trap mass measurements. In fact, the average deviation to the new masses determined in this work is 3$\sigma$, but extends to values as high as 6$\sigma$ for the neutron separation energies of $^{97}$Sr and $^{99}$Sr.

Fig.~\ref{fig:compare_nice} shows the resulting composition produced by the $r$-process for both cases. 
Although the masses used for this figure stem from AME03 \cite{Audi2003b} with AMEhigh and AMElow for the isotopes measured in this work, it clearly illustrates that the new masses introduce significant variations in the composition around $A=95-100$ due to their deviation from AME03 of up to 6$\sigma$ in neutron separation energies. 
The affected entropy components are about $S=70-110$. The component that is most dramatically affected by the new masses is the $S=100$ component, which is shown in Figs.~\ref{fig:nuclidcharta}, \ref{fig:compare_90a_nice}(a) and (b).
The reaction flows up to $A\approx90$ are characterized by a complex network of charged particle and neutron induced reactions and their inverse. This charged particle process provides the seeds for the $r$-process which then occurs at a somewhat later stage when temperatures have dropped and charged particle reactions have stopped to operate. The remaining free neutrons are then rapidly captured, driving the composition to more neutron-rich species and, via $\beta$ decays, up to heavier elements. The Sr isotopes are located in the transition region between these two types of reaction sequences and mark the lightest element involved in a ``rapid neutron capture'' reaction sequence at this entropy (at higher entropies neutron capture starts at lower element numbers, but the $r$-process converts all nuclei into heavier species so there is no longer a contribution to the $A\approx100$ mass region). 
Higher neutron separation energies for the Sr isotopes shift the reaction flow towards more neutron-rich nuclei. 
For low $S_n$ the dominant Sr waiting point, the point in an isotopic chain where the $\beta$ decay into the next isotopic chain occurs, is $^{98}$Sr. For high $S_n$, a significant fraction of the reaction flow proceeds via neutron capture on $^{98}$Sr to the $^{100}$Sr waiting point, leading to an increase in the production of $A=100$ nuclei. 

With the new Penning trap masses measured in this work, combined with the work of Hager et al. \cite{Hager2006}, the contribution of mass uncertainties in neutron-rich Rb and Sr isotopes to the $r$-process abundance pattern becomes negligible. This is shown in Fig.~\ref{fig:compare_90b_nice}(b) which shows essentially no change when varying the new neutron separation energies within their new 3$\sigma$ uncertainties, with NEWhigh being the variation up and NEWlow the variation down.

Also shown in Fig.~\ref{fig:compare_nice} are the solar $r$-process abundances \cite{Travaglio2004}, revealing the common problem of all models of this type in reproducing the solar composition of the light $r$-process elements with $A<115$. While it is apparent that the mass uncertainties of the nuclei considered here cannot explain this discrepancy, our measurements are a good step towards removing the nuclear physics uncertainties so one can better characterize the disagreement. 

A somewhat surprising result is that changes in the Rb and Sr masses in the $A=96-100$ range result in significant abundance changes across the entire $S=100$ component with significant changes occurring for mass numbers as low as $A=70$. Clearly the neutron capture reaction flow in the Sr region feeds back into the nucleosynthesis of the charged particle reaction sequence. The only possible explanation is that the switch in Sr waiting points affects the free neutron abundance,
which indeed is the case here. 
The shift of the reaction path towards more neutron-rich nuclei for AME03high leads to a reduction in the neutron abundance. Several conclusions can be drawn from this. Firstly, masses can affect the final $r$-process abundances globally, including the production of nuclei with lower mass number. Secondly, neutrons clearly play an important role in shaping the composition produced by the charged particle reaction sequence leading to an interplay between $r$-process and seed production.

\section{Conclusion and future improvements}
We have measured the masses of $^{94,97,98}$Rb and $^{94,97-99}$Sr to a precision better than 4~keV using the Penning-trap mass spectrometer TITAN. This presents mass measurements of highly charged ions in the charge state $q=15+$ and the first direct mass measurement of $^{98}$Rb. 
This work also influences the adopted mass values of several neighboring neutron-rich isotopes \cite{Audi2011}. 
Nuclear structure properties, such as the neutron separation energy S$_{2n}$, reveal and validate theory predictions of a sudden onset of large deformation from slightly deformed oblate or prolate shapes to strongly prolate shapes in the $58\leq~N\leq61$ region for Rb and Sr isotopes. This is manifested in Rb and more strongly in Sr with data presented in this work. 
In contrast, the more neutron-rich $61\leq~N\leq63$ region reveals no shape transition, and the smooth S$_{2n}$ trend is stabilized again. 
The precise and accurate mass measurements of this work 
will allow one to validate theoretical models and refine calculations towards more neutron-rich Rb and Sr. 
This may enable models such as the self-consistent mean-field approximation based on the D1S- or D1M-Gogny energy density functional (\cite{Rodriguez-Guzman2011} and references therein) to be tested under extreme conditions. 

The differences to previous work of up to 11$\sigma$ deviation in mass and 6$\sigma$ deviation in neutron separation energies motivated a study of the $r$-process. 
The $r$-process model calculations indicate that the measured masses are now known to a precision where their uncertainty does not contribute to model uncertainties anymore.

To strengthen the mass measurement program of HCI with TITAN, a preparation trap is presently being built. The goal is to increase the precision of mass measurements with HCI and to compensate for negative effects due to the charge-breeding process (efficiency losses, energy spread, etc.), and therefore a cooler Penning trap (CPET) will be commissioned \cite{Ryjkov2005, Simon2011, Simon2011a} in the near future. 
Further extension of mass measurements in this region is planned.

\section*{Acknowledgments}

This work has been supported by the Natural Sciences and Engineering Research Council of Canada. 
V.V.S. acknowledges support from the Studienstiftung des Deutschen Volkes and the German Academic Exchange Service (DAAD), T.B. from the Evangelisches Studienwerk e.V. Villigst, S.E. from the Vanier CGS program, and H.S. from NSF grants PHY06-06007 and PHY08-22648 (Joint Institute for Nuclear Astrophysics).



%

\end{document}